# Femtosecond laser inscription of nonlinear photonic circuits in Gallium Lanthanum Sulphide glass

María Ramos Vázquez[1,2], Belén Sotillo[3,4], Stefano Rampini[3], Vibhav Bharadwaj[3,5], Behrad Gholipour[6], Paloma Fernández[4], Roberta Ramponi[3], Cesare Soci[2] and Shane M. Eaton[3]

[1] ICRM, Interdisciplinary Graduate School, Nanyang Technological University, Singapore
[2] Centre for Disruptive Photonic Technologies, TPI, SPMS, Nanyang Technological University, Singapore
[3] Institute for Photonics and Nanotechnologies (IFN) -CNR, Milano, Italy
[4] Dept. Física de Materiales, Fac. Físicas, U. Complutense, Ciudad Universitaria, Madrid, Spain
[5] Center for Nano Science and Technology, Istituto Italiano di Tecnologia, Milano, Italy
[6] Optoelectronics Research Centre, University of Southampton, UK

E-mail: shane.eaton@gmail.com

## Abstract

We report on femtosecond laser writing of single mode optical waveguides in chalcogenide Gallium Lanthanum Sulphide (GLS) glass. A multiscan fabrication process was employed to create waveguides with symmetric single mode guidance and low insertion losses for the first time at 800 nm wavelength, compatible with Ti:Sapphire ultrafast lasers and nonlinear photonics applications. μRaman and X-Ray microanalysis were used to elucidate the origin of the laser-induced refractive index change in GLS. We found that the laser irradiation produced a gentle modification of the GLS network, altering the Ga-S and La-S bonds to induce an increase in the refractive index. Nonlinear refractive index measurements of the waveguides were performed by finding the optical switching parameters of a directional coupler, demonstrating that the nonlinear properties were preserved, evidencing that GLS is a promising platform for laser-written integrated nonlinear photonics.

Keywords: Ultrafast laser writing, chalcogenide glass, optical switching, Kerr effect, optical nonlinearities

## 1. Introduction

Nonlinear optical processes gained increasing attention in the past decade as the basis for innovative technologies in optics and photonics and in particular for all-optical signal processing [1]. The highly nonlinear properties of chalcogenide glasses, i.e. glasses made of one or more chalcogen elements (O, S, Se, Te, Po) and network formers (such as As, Ge, Sb, Ga, Si or P ), have attracted great interest as a nonlinear photonics platform: they have high linear and nonlinear refractive indices, a wide transparency window up to the infrared (IR) region of the spectrum, a low maximum phonon energy and an ultrafast nonlinear response time in the femtoseconds range [2]. Among the chalcogenides, commercially available Gallium Lanthanum Sulfide

(GLS) has the highest nonlinear figure of merit [3], being thermally stable up to 550°C, and does not present toxicity related issues as other commonly used chalcogenide glasses such as $As_2Se_3$.

Several approaches are available for creating optical circuits in chalcogenide glasses within thin films, microspheres or optical fibers [4]. Of particular interest is the fabrication of waveguides and other optical components via femtosecond laser inscription [5]. This technology relies on the nonlinear absorption of ultrashort laser pulses focused in the bulk of the material. The nonlinearity of the process leads to a localized modification which, in most glasses generates a positive refractive index change. Unlike planar silica [6] and silicon [7] technology, laser writing is a versatile technique which allows the formation of waveguides in the bulk of a wide variety of dielectrics [8, 9, 10, 11] in three dimensional geometries.

GLS waveguides with low propagation losses of 1.5 dB/cm at 633 nm wavelength [5] and 0.65 dB/cm at 1560 nm wavelength [12] have previously been demonstrated with femtosecond laser writing. In addition, femtosecond inscribed mid-IR photonic circuits for astronomical applications were reported with ~1 dB/cm loss at 3.39 µm wavelength [13, 14].

Here we report the femtosecond laser writing of buried waveguides in GLS designed for the convenient 800 nm band compatible with Ti:Sapphire technology for ultrafast nonlinear switching [15] and single photon sources for quantum information processing [16].

Waveguides were laser inscribed using the multiscan shaping approach, thus generating a gentle modification and avoiding nonlinear propagation effects to obtain a symmetric refractive index profile [13]. To gain further insight into the laser-formed modifications, combined analysis from micro-Raman (µRaman) spectroscopy and energy dispersion X-ray (EDX) characterization was performed for the first time in GLS. Furthermore, directional couplers were optically written into GLS and showed to work as nonlinear optical switches, with the nonlinear refractive index of the waveguides deduced from the optical switching parameters. The results presented here represent the first demonstration of laser-written photonic circuits in chalcogenides operating at the convenient 800 nm wavelength.

**2. Materials and methods**

*2.1 Glass synthesis*

Glasses were prepared by batching the constituent components in a controlled environment ($N_2$ atmosphere) glovebox. $Ga_2S_3$, $La_2S_3$ and $La_2O_3$ were batched and mixed thoroughly to obtain a homogeneous mixture. Vitreous carbon crucibles used for the melt were washed with de-ionized (DI) water and put in an ultrasonic bath of DI water for 1 h, then immediately put into vacuum oven at a temperature of 80°C overnight.

Melting was carried out in an in-house designed horizontal tube furnace equipped with an adjacent water-cooled jacket, in order to get adequate quench rates to obtain a glassy melt. The furnace temperature was raised with a ramp rate of 20°C/min to 1150°C and held for 24 h. The melt was subsequently pushed out into the aforementioned water jacket, where it rapidly cooled from 1150°C to room temperature. The glass was annealed at 530°C for 24 h in order to relieve any stresses inherent as a result of the quenching process. The linear refractive index of the GLS glass is 2.4 at 800 nm wavelength.

*2.2 Femtosecond laser inscription of GLS waveguides*

The femtosecond laser used for waveguide writing in GLS was a regeneratively amplified Yb:KGW system (Pharos, Light Conversion) with 230 fs pulse duration, 1030 nm wavelength, focused with a 0.42 NA microscope objective (M Plan Apo SL50X Ultra-Long Working Distance Plan-Apochromat, Mitutoyo). The repetition rate of the laser can be varied from 1 MHz to single pulse, but for the laser writing experiments was held fixed at 500 kHz. Computer-controlled, 3-axis motion stages (ABL-1000, Aerotech) interfaced by CAD-based software (ScaBase, Altechna) with an integrated acousto-optic modulator were used to translate the sample relative to the laser to form the desired photonic structures. The waveguides were scanned transversally with respect to the incident laser, with the polarization perpendicular to the scan direction.

Waveguides were inscribed at a depth of 150 µm below the glass surface, with incident laser power *P* ranging from 10 mW to 250 mW (laser pulse energies from 20 nJ to 500 nJ). Each waveguide consisted of 18 consecutive laser scans separated by 0.3 µm. Four translation speeds *v* = 5, 10, 20, and 50 mm/s were used to inscribe each batch of waveguides. The sample facets were polished after the inscription for an optimum optical quality. Sets of directional couplers were fabricated with an input-to-input port distance *p* = 100 µm, *R* = 45 mm bend radius and a center to center separation distance in the interaction region of *s* = 7 µm. The interaction length *L* was varied between 0 and 2 mm in steps of 0.25 mm.

*2.3 Waveguide and directional coupler characterization*

Waveguide transmission measurements were performed using butt-coupled fibers. High resolution 3-axis manual positioners (Nanomax MAX313D, Thorlabs) were used for fiber in and fiber out alignments. The four-axis central waveguide manipulator (MicroBlock MBT401D, Thorlabs) enabled transverse displacement between sets of waveguides in the sample. A light source at 808 nm (S1FC808, Thorlabs) was coupled to the waveguides using single-mode fiber (780HP, Thorlabs). At the output, light was fiber-coupled to an optical power meter (818-SL, Newport) to measure the power transmitted through the waveguide. To measure the near-field waveguide mode profile, a 60× asphere (5721-H-B, Newport) was used to image the light onto a beam profiler (SP620U, Spiricon).

*2.3 Nonlinear characterization of directional couplers*

The nonlinear characterization of the directional couplers was performed with a home built amplified Ti:Sapphire laser system operating at 2 kHz repetition rate, 180 fs pulse width and 800 nm wavelength was used to couple the light into one of the input ports of the directional couplers by focusing the light with a 0.16-NA antireflection coated objective. A 0.25-NA microscope objective was used to collect the output guided modes from the end facet of the glass sample and subsequently light was sent into a Si CCD camera for power detection (SP620U, Spiricon). The light power at the input was measured using a silicon photodetector placed before the focusing objective.

*2.4 Raman spectroscopy*

To gain better understanding of the waveguide formation in bulk GLS, µRaman measurements were performed. The µRaman spectra were recorded in a Horiba Jovin Yvon LabRAM HR800 confocal microscope, equipped with an air-cooled CCD. The excitation wavelength was 633 nm from a He-Ne laser source. A 100× (0.9 NA) objective was used to focus the laser on the sample as well as to collect the Raman spectra (backscattering configuration), with a spatial resolution below 1 µm.

*2.5 EDX compositional analysis*

EDX was performed to study the compositional element analysis of the laser formed waveguides. EDX measurements were done in a Bruker Quantax 70 EDS system attached to a scanning electron microscope Hitachi TM3000, operated at 15 kV. The energy resolution for this system is around 154 eV (for Cu Kα). Backscattered electron (BSE) images were taken also in the Hitachi TM300 SEM.

**3. Results and discussion**

*3.1 Waveguide characterization*

Figure 1 shows the cross-sectional (a) and overhead (b) microscopic view and mode profile (c) for the lowest loss waveguide, formed with a pulse energy of 70 nJ and a scan speed of 5 mm/s. For energies above 80 nJ, the waveguides were multimode. The insertion loss of the optimum waveguide was 2.0 dB and the coupling loss was estimated to be 0.7 dB/facet, obtained from the overlap integral between the waveguide (mode field diameter, MFD = 7.5 µm) and fiber (MFD = 5 µm) modes. After accounting for a Fresnel reflection of 0.2 dB/facet, we infer a propagation loss of ~0.1 dB/cm. Assuming a step profile of the refractive index change and given the experimentally measured MFD of 7.5 µm, we estimate a refractive index increase of $\Delta n = 10^{-3}$ using a commercial mode solver (Lumerical MODE Solutions). Due to the gentle and multiscan laser writing used to form the waveguide cross section, it is a reasonable assumption that the refractive index profile follows a step profile. This is in contrast to a higher pulse energy and single scan approach, which leads to a more complex refractive index distribution.

The slightly rectangular waveguide cross section shown in Fig. 1(a) formed with the multiscan method led to the best compromise between a tightly confined mode and low propagation loss. When using more transverse scans to produce a square waveguide cross section, the waveguide became multimode.

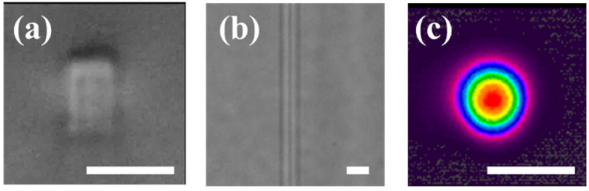

**Fig. 1.** Transverse (a) and overhead (b) microscope view of the waveguide written with 70 nJ and 5 mm/s. (c) The guided mode at 808 nm wavelength. The scale bar corresponds to 10 μm.

### 3.2 Directional couplers characterization

As shown in Fig. 2, we defined directional couplers with S-bends which bring two waveguides in close proximity to enable evanescent light coupling, and with the following geometry: radius of curvature $R$, interaction length $L$, center to center separation in the interaction region $s$, separation between the two ports $p$.

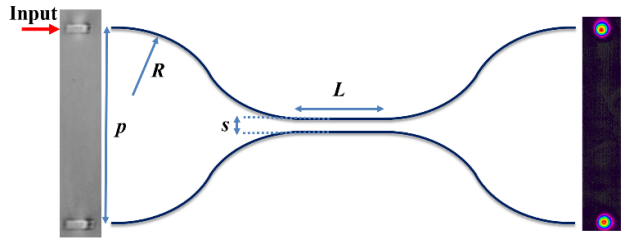

**Fig. 2.** Sketch of a symmetric directional coupler showing the various parameters defining it radius of curvature $R$, interaction length $L$, center to center separation between the waveguides in the interaction region $s$, separation between the two ports $p$. On the left, a side view optical microscope image of the two input ports is shown and, on the right, the output modes measured for a directional coupler with 50% coupling ratio is shown.

To find the optimum radius of curvature for the S-bends, a study of bend loss with varying radius of curvature was performed. The bend loss was calculated by subtracting the insertion loss of the straight waveguide from the insertion loss of the two S-bends making up one arm of the directional coupler. Figure 3 shows the bend loss versus radius of curvature. As expected, larger bend radii offer lower losses, but result in longer axial length of the S-bends, increasing the overall size of the chip (defined as axial footprint of a directional coupler with zero interaction length). The best compromise between the losses and the size of the chip was obtained with a radius of curvature of 45 mm.

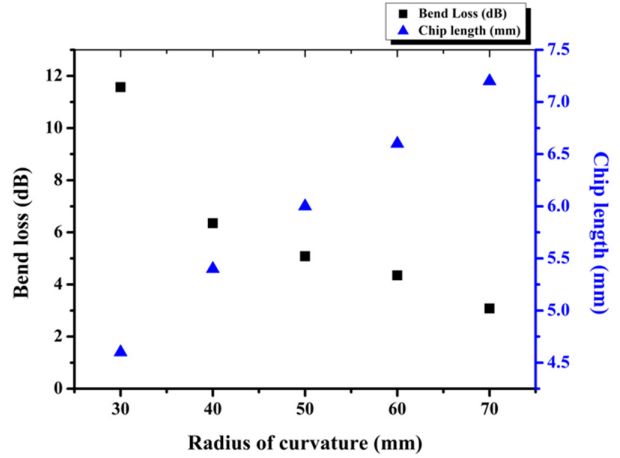

**Fig. 3.** Measured bend loss of two S bends and chip length (two S bends with zero interaction length) versus radius of curvature.

Symmetric directional couplers were laser written in the bulk of the GLS sample with varying interaction length $L$ for $R$ = 45 mm and separation between the waveguides in the interaction region $s$ = 7 μm. The optimum parameters (pulse energy 70 nJ and scan speed 5 mm/s) for straight waveguide fabrication were used to fabricate the couplers. The separation between the ports $p$ was fixed at 100 μm, a convenient separation for fiber characterization.

The laser written directional couplers were characterized by fiber launching 808 nm wavelength light into one arm of the directional coupler and measuring the power at the output of the cross and through output ports. A plot of coupling ratio versus the interaction length was obtained, as shown in Fig. 4. This plot was well fitted to a sine square function according to the equation $\mathrm{CR} = \sin^2\left(\frac{\pi}{l_B(\lambda)}L\right)$, where CR is the power coupling ratio, $l_B$ the coupler beat length, $\lambda$ the wavelength and $L$ the interaction length. The peak coupling ratio obtainable was 54%, somewhat short of the ideal 100%, due to differing propagation constants of the two waveguides forming the directional coupler [17]. As the waveguides were fabricated in the non-heat accumulation regime with multiple passes, the waveguides are more sensitive to any fluctuations during fabrication compared to waveguides written with the cumulative heating process [17]. Further optimization of the laser-written waveguides could lead to an increase in this peak coupling ratio.

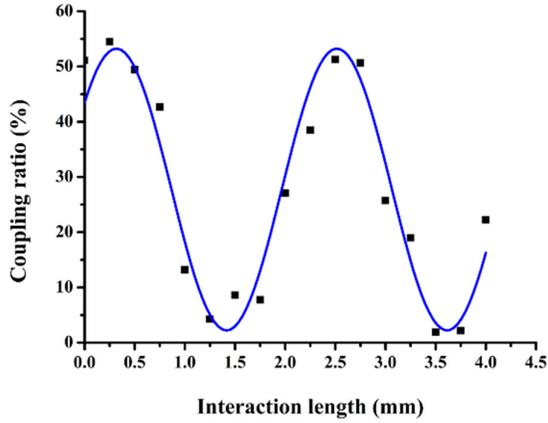

**Fig. 4.** Plot of coupling ratio with respect to the interaction length with the sin squared fitting represented in blue.

*3.3 Nonlinear switching in directional couplers*

An all-optical switch is a device that allows the control of an optical signal through the use of another optical signal [15]. In optical communication systems, large bandwidths, ultrafast switching speeds below the sub-picosecond range and processing signals already in the form of light can be especially profitable. Specifically, researchers look for optical devices that fulfil requirements for practical applications in terms of low power consumption, low switching thresholds, fast switching times and high transmission ratios. Some previous key example devices working as all-optical switches are photonic crystals, metamaterials, microring resonators and surface plasmon polaritons [15]. Among all of them, third order nonlinear directional couplers present the fastest switching speeds, at the femtosecond level, but with the disadvantage of owning lower transmittance ratios and slightly higher pump power thresholds than their counterparts [15].

The nonlinear directional couplers act as an ultrafast all-optical switching device due to the power control of a unique input signal. In this case, the foundation of all-optical switching relies on the refractive index change of the material, induced by the nonlinear Kerr effect. Therefore, all-optical switching is manifested as an intensity dependent refractive index change induced in the nonlinear material [18]:

$$n = n_0 + n_2 I,$$

where $n_0$ is the linear refractive index, $n_2$ is the nonlinear refractive index and $I$ is the intensity of the control light.

When an intense electromagnetic field propagates through the glass waveguide, a nonlinear polarization manifests itself in the medium and subsequently modifies the propagation properties of the light. In the case of GLS glass, this nonlinear effect occurs because of the high third order hyperpolarizability of the sulphide ion [4].

Therefore, the light propagation through a directional coupler will periodically alternate from one waveguide to the other one while the input irradiance changes. At low input irradiances, the propagation of light will remain in the linear regime. At high input irradiances, the nonlinear effect induced in the medium will lead to self-focusing or self-trapping causing most of the light to remain in the through output port waveguide [18].

The directional coupler selected for the measurement of the nonlinear refractive index had an interaction length $L$ = 250 µm for ~50/50 coupling in the linear regime. The relative transmission of the through and cross output waveguide ports as a function of the input irradiance is shown in Fig. 5. The switching input irradiance needed to achieve the same power in the through (bar) and cross ports was $7.1 \times 10^{15}$ W/m². For lower input irradiances, most of the light was evanescently coupled into the cross waveguide, with a relative transmission of the cross output port at 54%. For values above the switching irradiance, most of the light is in the through port, with the relative transmission of the cross port decreasing to 48%. We observed that this process was fully reversible by lowering the input power.

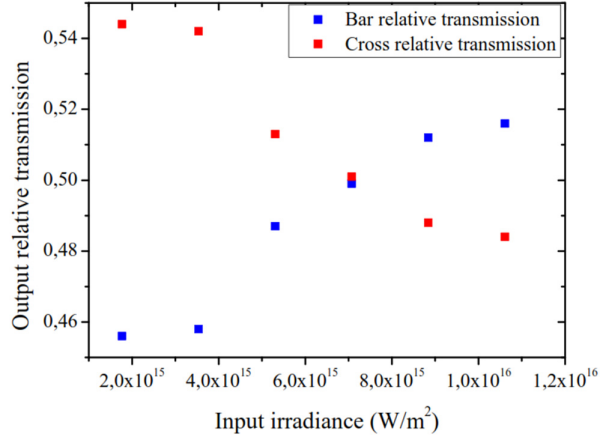

**Fig. 5.** Bar and cross output relative transmissions as a function of input irradiance. The point at which the output relative transmissions become equal is for an input irradiance of 7.1 × 10$^{15}$ W/cm$^2$.

The nonlinear phase change induced within the directional coupler, due to nonlinear refraction, is given by [19]:

$$\Delta\phi = \frac{2\pi L n_2 I}{\lambda},$$

where $L$ is the interaction length of the directional coupler, $n_2$ is the nonlinear refractive index of the waveguide, $I$ is the input irradiance and $\lambda$ is the operating wavelength.

Taking into account that the phase change required to switch a nonlinear directional coupler is approximately 4π [19] and considering the Fresnel losses at the input facet of the waveguide, an evaluation of the nonlinear refractive index of the laser written waveguides can be performed. For the directional coupler, the nonlinear refractive index of the waveguide is 9.0 × 10$^{-19}$ m$^2$/W, on the same order of magnitude as the value reported for pristine GLS [4]. Kar's group previously reported a reduction in the nonlinear refractive index for laser-written waveguides in GLS when characterized at 1550 nm [20]. We attribute this discrepancy due to the gentler modification we used to form waveguides with a lower refractive index change for the single mode guiding of 800 nm light.

3.4 Structural and compositional characterization

Initial characterization of the pristine glass has been performed to understand the changes that the femtosecond laser writing is producing in the material. In Fig. 6(a) the absorption spectrum of a sample is shown, where the absorption edge is around 515 nm, similar to previously reported values [21]. An EDX spectrum is presented in Fig. 6(b), confirming the presence of S, Ga, and La in a proportion 4:2:1. EDX measured at different points and in different samples indicate a uniform distribution of the three elements. No oxygen signal was detected.

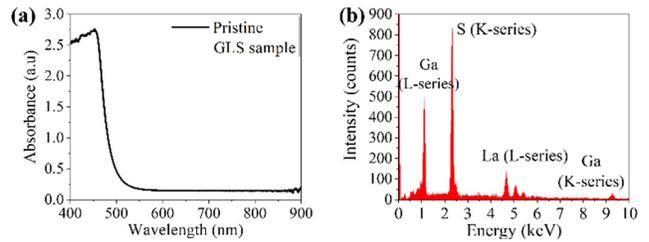

**Fig. 6.** (a) Absorbance and (b) EDX spectra of pristine GLS.

Before explaining the Raman spectra recorded in GLS glass, a brief description of the glass network is needed to understand the spectra. Ga atoms are bonded to sulfur forming tetrahedral networks of GaS$_4$, which are the glass units. S$^{2-}$ anions provided by the modifying lanthanum sulfide maintain the tetrahedral environment of GaS$_4$, as gallium sulfide alone cannot form a glass. The negatively charged GaS$_4$ tetrahedra are reception sites for La$^{3+}$ cations [22].

The Raman spectra of the GLS glass is then generally dominated by the modes associated with the GaS$_4$ tetrahedra vibrations [23, 24]. As shown Fig. 7, the main band is located between 250 and 450 cm$^{-1}$ and centred at 325 cm$^{-1}$. Other bands are also visible in the region 100 – 250 cm$^{-1}$. The dominant band at 325 cm$^{-1}$ is ascribed to symmetrical stretching vibrations of the GaS$_4$ units, whereas the band at 140 cm$^{-1}$ is associated to the asymmetrical bending. Band around 100 cm$^{-1}$ and 380 cm$^{-1}$, which are less defined in the spectrum, come from the symmetrical bending and the asymmetrical bending, respectively. The band related to the La-S vibration is found at 215 cm$^{-1}$ [25]. There is no evidence of crystallization in the Raman spectra taken in the pristine material as well as in the spectra of the waveguides.

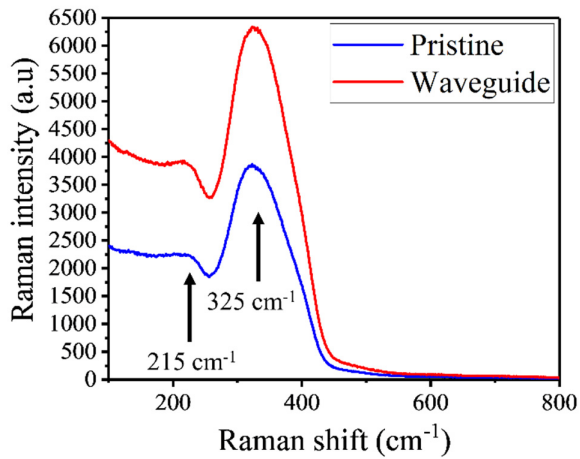

**Fig. 7.** μRaman spectra recorded on pristine GLS (blue curve) and on a waveguide (red curve) written with a laser energy of 70 nJ and a scan speed of 5 mm/s. The excitation wavelength is 633 nm.

In the μRaman spectra recorded inside the waveguides (Fig. 8), there are two main effects observed. The first effect is the increase of the total Raman signal (Fig. 7 and 8(a)). The second observation is the increase of the relative intensity of the band associated with La-S bonds at 215 cm$^{-1}$ (Fig. 8(b)-(c)). Also, slight changes of relative intensity of the different vibration modes of the GaS$_4$ tetrahedra are observed. These variations of the μRaman spectra are uniform inside the region of the waveguide, as shown in the maps of Fig. 8(a)-(c).

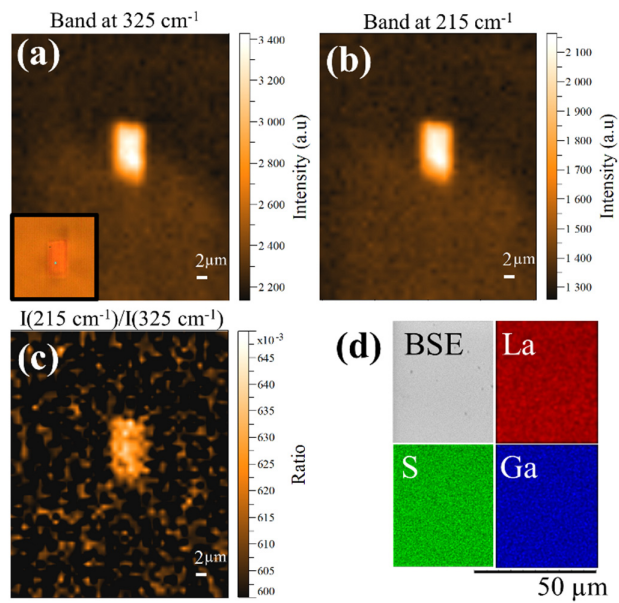

**Fig. 8.** (a)-(c) μRaman maps of different bands in a waveguide written with a laser energy of 70 nJ and a scan speed of 5 mm/s. (a) Intensity of the band at 325 cm$^{-1}$ (symmetrical stretching vibrations of the GaS$_4$ units). The inset shows the transverse optical microscope image of the waveguide. (b) Intensity of the band at 215 cm$^{-1}$ (La-S bond vibrations). (c) Intensity ratio of the 215 cm$^{-1}$ and 325 cm$^{-1}$ band. All the scale bars are 2 μm. (d) BSE image and EDX maps of the different elements (La in red, S in green, Ga in blue).

The changes in the relative intensity of the GaS$_4$ related modes is related to slight deformations of the bonds of the tetrahedral (i.e. bonding angles). Also the increase of the total Raman intensity is related to a deformation of bonds in the GLS network, which produces an increase of polarizability and/or density of the material in the waveguide. This produces the observed increase in the refractive index. The increase of the relative intensity of the La-S vibration band is related to the breaking of Ga-S bonds and the formation of new La-S bonds.

In other results of femtosecond laser writing of waveguides in glasses containing lanthanum, the increase in the refractive index was attributed to ion migration [25, 26]. For the GLS system utilized in this work, and for all the waveguides studied, no Z-contrast was observed in the BSE images, which suggests that the laser did not produce a strong elemental compositional change in the irradiated glass network. It is possible that the heat accumulation regime during femtosecond laser writing [27, 28] could lead to a migration of ions in GLS. From the EDX measurements, presented in Fig. 8(d) (La in red, S in green and Ga in blue) no significant migration of the elements is observed inside the waveguide. These observations are in agreement with previously reported EDX studies performed in GLS waveguides, where no compositional changes were detected [29]. Then, from EDX and Raman measurements, the laser is inducing changes of the Ga-S and La-S bondings in the focal volume, producing a rearrangement of the glass network that leads to an increase in the refractive index.

If the pulse energy used to fabricate the waveguides is too high, dark filaments appear around the waveguide. The Raman signal in these darker region decreases (see Fig. 9(a)-(b)), an effect that is typically associated with extended defects formation and broken bonds [12]. So if the pulse energy is too high, the glass is not able to rearrange the bonds and the glass network suffers an expansion (evidenced also in the BSE images with a swollen region at the surface, not shown) and a decrease of the refractive index. This is the opposite effect to the one observed for low pulse energies. Further increment of the laser power produces a high damage in the glass, expanding the glass network and creating voids that sometimes emerge to the surface, as shown in the BSE image of Fig. 9(c). EDX map of Fig. 9(d) shows that the material (mainly Ga and S) is ejected outside the laser-written tracks. The creation of voids for higher pulse energies has also been observed in other glasses like fused silica [30].

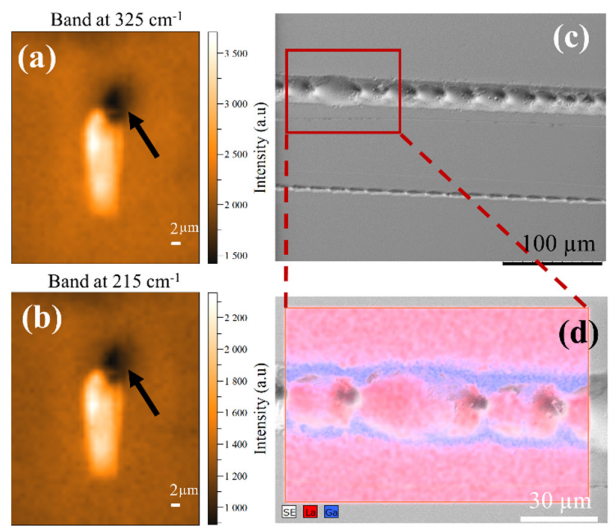

Fig. 9. (a)-(b) μRaman maps of different bands in a waveguide written with 200 nJ and 10 mm/s. (a) Intensity of the band at 325 cm$^{-1}$. (b) Intensity of the band at 215 cm$^{-1}$. (c) BSE image of a track fabricated with 400 nJ. The compositional map of a selected area is shown in (d) (La red, Ga blue).

## 4. Summary

In summary, we fabricated waveguides in GLS using focused femtosecond laser pulses, demonstrating single mode guidance and an insertion loss of 1.9 dB at 800 nm wavelength. Ultrafast all-optical switching was performed with femtosecond laser pulses using the laser written directional couplers and the nonlinear refractive index was found to be on the same order of magnitude as the pristine material. EDX and Raman measurements performed in the waveguides showed that the laser irradiation produced a gentle modification of the GLS network, altering the Ga-S and La-S bonds to induce an increase in the refractive index.

**Acknowledgements**

We are grateful to Luigino Criante for access to the FemtoFab facility at CNST–IIT Milano for the laser fabrication experiments. We appreciate the funding sources of Singapore Ministry of Education (MOE2016-T3-1-006), DIAMANTE MIUR-SIR Grant and FemtoDiamante Cariplo ERC Reinforcement Grant. B. Sotillo acknowledges financial support from Comunidad de Madrid (Ayudas atracción de talento) and Spanish Ministry Economy, Industry and Competitiveness (MINECO/FEDER-MAT2015-65274-R Project). The authors would like to thank Prof. Daniel Hewak and Mr. Christopher Craig for help with glass synthesis. The University of Southampton acknowledges the support of the Engineering and Physical Sciences Research Council through the EPSRC Centre for Innovative Manufacturing in Photonics (EP/H02607X/1). The data for this is accessible through the University of Southampton Institutional Research Repository (doi: XXXX).